\documentclass[floatfix,aps,prd,nofootinbib,superscriptaddress,preprint]{revtex4-1}

\usepackage{amsmath,amssymb,bm,bbm,amsfonts}
\usepackage{physics}
\usepackage{graphicx,graphics}
\usepackage[dvipsnames]{xcolor}
\usepackage[colorlinks=true,linkcolor=blue,citecolor=blue,urlcolor=blue]{hyperref}
\usepackage{dsfont}
\usepackage{comment}
\usepackage{hhline,multirow}
\usepackage{dcolumn}
\usepackage{url}
\usepackage[normalem]{ulem}
\usepackage{mathrsfs}
\usepackage{latexsym}
\usepackage{amscd}
\usepackage{slashed}
\usepackage{mathtools}

\newcommand{\beq}{\begin{Eqarray}}
\newcommand{\eeq}{\end{eqnarray}}
\newcommand{\beqs}{\begin{eqnarray*}}
\newcommand{\eeqs}{\end{eqnarray*}}

\newcommand{\true}{{\mathrm{true}}}
\renewcommand{\min}{{\mathrm{min}}}

\newcommand{\vc}[1]{{\bm #1}}

\begin{document}

\preprint{JLAB-THY-22-3638, \, ADP-22-2/T1173}

\title{\mbox{On the determination of uncertainties in parton densities}}

\author{N.~T.~Hunt-Smith}
\affiliation{CSSM and ARC Centre of Excellence for Dark Matter Particle Physics, Department of Physics, The University of Adelaide, Adelaide 5005, Australia}
\author{A.~Accardi}
\affiliation{Hampton University, Hampton, Virginia 23668, USA}
\affiliation{Jefferson Lab, Newport News, Virginia 23606, USA}
\author{W.~Melnitchouk}
\affiliation{Jefferson Lab, Newport News, Virginia 23606, USA}
\author{N.~Sato}
\affiliation{Jefferson Lab, Newport News, Virginia 23606, USA}
\author{A.~W.~Thomas}
\affiliation{CSSM and ARC Centre of Excellence for Dark Matter Particle Physics, Department of Physics, The University of Adelaide, Adelaide 5005, Australia}
\author{M.~J.~White}
\affiliation{CSSM and ARC Centre of Excellence for Dark Matter Particle Physics, Department of Physics, The University of Adelaide, Adelaide 5005, Australia}

\begin{abstract}
We review various methods used to estimate uncertainties in quantum correlation functions, such as parton distribution functions (PDFs).
Using a toy model of a PDF, we compare the uncertainty estimates yielded by the traditional Hessian and data resampling methods, as well as from explicitly Bayesian analyses using nested sampling or hybrid Markov chain Monte Carlo techniques.
We investigate how uncertainty bands derived from neural network approaches depend on details of the network training, and how they compare to the uncertainties obtained from more traditional methods with a specific underlying parametrization.
Our results show that utilizing a neural network on a simplified example of PDF data has the potential to inflate uncertainties, in part due to the cross validation procedure that is generally used to avoid overfitting data.

\end{abstract}

\date{\today}
\maketitle

%%%%%%%%%%%%%%%%%%%%%%%%%%%%%%%%%%%%%%%%%%%%%%%%%%%%%%%%%%%%%%%%%%%%%%%%%%%
\section{Introduction}
\label{sec:introduction}

The partonic structure of hadrons plays a fundamental role in elementary particle and nuclear physics.
Understanding the spin structure of the nucleon and the hadronization of partons, interpreting experimental data according to the Standard Model (SM) of fundamental interactions, precisely measuring SM parameters and searching for signals of physics beyond the SM, are only some of the phenomenological problems that rely on knowledge of parton distribution functions (PDFs)~\cite{Jimenez-Delgado:2013sma, Gao:2017yyd, Ethier:2020way, Kribs:2020vyk, Thomas:2021lub} and fragmentation functions (FFs)~\cite{Metz:2016swz}.
Because these functions cannot yet be computed sufficiently accurately from quantum chromodynamics (QCD), both PDFs and FFs are determined by analysing experimental data from a wide range of hard-scattering processes within the framework of collinear factorization~\cite{Collins:1989gx}.
Here, physical observables are factorized into hard-scattering partonic cross sections, calculable perturbatively from QCD, and nonperturbative distribution functions, whose evolution with the momentum scale probed in the measured process can nonetheless be calculated perturbatively \cite{Gribov:1972ri, *Lipatov:1974qm, *Altarelli:1977zs, *Dokshitzer:1977sg}.

Considerable progress has been made in recent years through several parallel efforts to improve our knowledge of PDFs, both experimentally and theoretically (for simplicity in the following we will refer only to PDFs, but the discussion applies equally to FFs).
The inclusion of an increasingly vast and precise dataset and the growth of the theoretical accuracy of the QCD analyses have been accompanied by an increasing sophistication of the methodological procedures used to represent the uncertainties on the PDFs \cite{Kovarik:2019xvh}.
These currently include the Hessian~\cite{Pumplin:2001ct}, Lagrange multiplier~\cite{Stump:2001gu} and data resampling (DR) methods. 
The uncertainty estimation method most commonly adopted in the literature has been the Hessian, which has been used by many collaborations in their determination of unpolarized PDFs (see Refs.~\cite{Hou:2019efy,Bailey:2020ooq,Gao:2017yyd} for a review) and by the DSS group in their determination of the FFs of light hadrons~\cite{deFlorian:2014xna, deFlorian:2017lwf}.
The Lagrange multiplier method was used by the DSSV group to determine the  polarized proton PDFs~\cite{deFlorian:2008mr, *deFlorian:2009vb, deFlorian:2014yva}, while the DR method was used by the NNPDF collaboration~\cite{Forte:2002fg, *DelDebbio:2004xtd, *DelDebbio:2007ee, *Ball:2008by} in their determination of the unpolarized and polarized proton PDFs (for recent work, see, {\it e.g.}, Refs.~\cite{NNPDF:2021njg, Nocera:2014gqa}) and of the parton to light hadron FFs~\cite{Bertone:2017tyb}.
It was also employed by the JAM collaboration in their determinations of the polarized proton PDFs~\cite{Sato:2016tuz} and the pion and kaon FFs~\cite{Sato:2016wqj}, and for the first simultaneous fit of these two~\cite{Ethier:2017zbq}, as well as more recently in the JAM global QCD analyses of unpolarized proton \cite{Cocuzza:2021rfn, Cocuzza:2021cbi} and pion \cite{Cao:2021aci, Barry:2021osv} PDFs.
Although the NNPDF and JAM approaches share the same method of quantifying uncertainties, they differ in their parametrization of the PDFs.
While the JAM collaboration uses a traditional polynomial functional form for the dependence of the PDFs on the quark momentum fraction $x$, the NNPDF collaboration implements a similar basic parametric form supplemented by a series of trained neural network weights. 

We should note that the above methods all attempt to find approximations to the Bayesian posterior.
An alternative approach is to calculate the posterior exactly using Bayes' theorem.
This can then be used to obtain the expectation value and standard deviation of the set of PDF parameters, and of any derived quantity.
A popular posterior mapping technique is nested sampling (NS)~\cite{skilling2006}, a Monte Carlo method targeted at the efficient calculation of the Bayesian evidence, which is the average likelihood of a model over its probability space. 
This method produces posterior inferences as an advantageous byproduct, and contemporary implementations of NS algorithms, such as 
    {\tt MultiNest}~\cite{Feroz:2008, *Feroz:2009} and 
    {\tt PolyChord}~\cite{Handley:2015, higson2017},
are now extensively used for parameter estimation from posterior samples in cosmology (see, {\it e.g.}~Ref.~\cite{refId0}).
They have also very recently been used in global QCD analyses of the quark transversity distributions~\cite{Lin:2017stx} and of the pion distributions~\cite{Barry:2018ort} by the JAM collaboration.

A popular alternative to nested sampling is the Hamiltonian (or hybrid) Markov chain Monte Carlo (HMC) technique~\cite{Duane:1987de, brooks2011handbook}, based on the Metropolis-Hastings algorithm~\cite{Metropolis:1953am, Hastings:1970aa}, which allows for an efficient sampling of posterior inferences, though without the calculation of the evidence which is typically not required in PDF fits.
These methods are implemented in xFitter~\cite{Alekhin:2014irh}, the open-source package for performing global PDF fits, which has been shown to lead to competitive results~\cite{Gbedo:2017eyp}.

It is evident that a reliable and faithful representation of PDF uncertainties is especially critical in applications such as searches for signals of physics beyond the SM, where sub-percent precision is often sought.
As the above survey suggests, it is difficult to compare uncertainties between groups using different methodologies in a meaningful and quantitative way~\cite{Ball:2022hsh, Butterworth:2015oua, Accardi:2016ndt}. 
This is especially challenging when there exist tensions between data sets, or when there are strong departures from the Gaussian approximation to probability distributions of parameters.

In this paper, we review and compare existing methods for estimating uncertainties in global QCD analyses of PDFs. 
Specifically, we use a computationally inexpensive ``toy'' model of a PDF fit to compare the uncertainty estimates that arise from the DR, Hessian, NS and HMC methods, and explore whether they produce the same results.
In each case, we assume that the functional form of the PDF being fitted is the same as that used to generate the toy data, so as not to introduce any inflated uncertainty arising from inaccuracies in the representation of the fitted functional form.

Furthermore, we investigate the case where the known functional form in the PDF fit is replaced by a neural network, and investigate how the uncertainty estimates that result from DR depend on the the cross validation procedure used in the training of the neural network.
We also compare the uncertainty estimates from the neural network approach to the other methods in a similar vein to a ``closure test''~\cite{Ball:2014uwa}, to establish how often the true values of our toy model fall within the uncertainty bands predicted by each method across many fits.

In Sec.~\ref{sec:formalism} we begin by reviewing the Hessian, DR, NS and HMC approaches to estimating PDF uncertainties.
Section~\ref{sec:comparison} describes the toy model, which is designed to reproduce the salient features of a real PDF fit, but with a significantly reduced computation time. 
We then perform a systematic comparison of the results obtained by applying
the Hessian, DR, NS and HMC techniques to the toy model analysis.
The study of the neural network approach is performed in Sec.~\ref{sec:RPCV}, where we in particular explore the effects of partitioning and cross validation in the training of neural networks on PDF uncertainties.
Finally, in Sec.~\ref{sec:conclusion} we present our conclusions and discuss future applications of this work.

%%%%%%%%%%%%%%%%%%%%%%%%%%%%%%%%%%%%%%%%%%%%%%%%%%%%%%%%%%%%%%%%%%%%%%%%%%%%
\section{Formalism}
\label{sec:formalism}

In the framework of Bayesian inference~\cite{DAgostini:1995jqe, 0034-4885-66-9-201, cowan1998statistical} one seeks to determine the probability measure $p([f])$ associated with a space of functions $[f]$.
The expectation value $E\{\mathcal{O}[f]\}$ and the variance $V\{\mathcal{O}[f]\}$ of any physical observable $\mathcal{O}$ that depends on a set of PDFs $[f]$ can then be expressed by integration over the space of functions, respectively, of the first and second statistical moments of $\mathcal{O}[f]$, evaluated on each function $f$ and weighted by the probability measure $p([f])$.

The probability measure {$p([f])$} is an infinite-dimensional object in the space of functions $[f]$.
However, it should be necessarily determined from a finite set of measurements.
The problem, which would otherwise be ill-defined, is usually projected from the infinite-dimensional space of functions into a finite-dimensional space of parameters by choosing a suitable parametrization for the set of functions $[f]$.
Such a procedure of course requires some assumptions.
Some of these, such as a certain degree of smoothness or particular behaviors in certain regions of the functional space, may be physically motivated; nevertheless, they should not bias the result nor affect its statistical  interpretation.

Calling $\vc{a}=\{a_1, \dots, a_{n_{\rm par}}\}$ the set of $n_{\rm par}$ parameters and $\vc{m}=\{m_1, \dots, m_{n_{\rm dat}}\}$ the set of $n_{\rm dat}$ measurements, the expectation value and variance of an observable $\mathcal{O}$ are defined in the Bayesian framework by
\begin{subequations}
\label{eq:EVbayes}
\begin{eqnarray}
E_{\rm Bayes}\{\mathcal{O}(\vc{a})\}
& =
\langle\mathcal{O}(\vc{a})\rangle
& =
\int\!\dd{\vc{a}}\,p(\vc{a}|\vc{m})\,\mathcal{O}(\vc{a})
\label{eq:expvaluepar}
\,,
\\
V_{\rm Bayes}\{\mathcal{O}(\vc{a})\}
& =
\sigma^2_{\mathcal{O}}(\vc{a})
& =
\int\!\dd{\vc{a}}\,p(\vc{a}|\vc{m})
\big(
\mathcal{O}(\vc{a}) - E_{\rm Bayes}\{\mathcal{O}(\vc{a})\}
\big)^2
\,,
\label{eq:variancepar}
\end{eqnarray}
\end{subequations}
where $p(\vc{a}|\vc{m})$ is the conditional probability of the set of parameters $\vc{a}$, given the set of measurements $\vc{m}$.
The problem of determining the probability measure in the space of functions $p([f])$ is then reduced to the problem of determining the conditional probability $p(\vc{a}|\vc{m})$.

One can readily formulate the problem in terms of Bayes' theorem, which can be expressed as
\begin{equation}
p(\vc{a}|\vc{m})
=
\frac{1}{\mathcal{Z}}\,p(\vc{m}|\vc{a})\,p(\vc{a})
\,,
\quad
\text{with}
\quad
\mathcal{Z} 
=
\int \dd{\vc{a}}\, p(\vc{m}|\vc{a})\,p(\vc{a})
\,,
\label{eq:Bayestheorem}
\end{equation}
where $p(\vc{a})$ is the prior, $p(\vc{a}|\vc{m})$ is the posterior, $p(\vc{m}|\vc{a}) \equiv \mathcal{L}(\vc{a};\vc{m})$ is the likelihood, and $\mathcal{Z}$ is the evidence.
The prior is a subjective quantity, and results of a Bayesian analysis are frequently found to be highly dependent on the choice of this function.
For problems where the data are sufficiently constraining (as expected in the case of PDF fits), this is less of an issue, and the posterior distribution is mostly determined by the likelihood function.
For linear problems, flat priors can be useful: $p(\vc{a})$ is associated with a uniform distribution, normalized to unit volume, $p(\vc{a})=\mathbf{1}$; Eq.~\eqref{eq:Bayestheorem} then implies that
\begin{equation}
p(\vc{a}|\vc{m})
\propto
\mathcal{L}(\vc{a};\vc{m})
\,.
\label{eq:flatprior}
\end{equation}

The likelihood $\mathcal{L}(\vc{a};\vc{m})$ is seen as a mathematical function of the parameters $\vc{a}$, given the data $\vc{m}$. 
For $n_{\rm dat}$ pairs of independent measurements $\vc{m}=\{x_i,y_i\}$, $i=1,\dots,n_{\rm dat}$, a common choice is
\begin{equation}
\mathcal{L}(\vc{a};\vc{m})
=
\mathcal{N}
\exp\Big[-\frac{1}{2}\chi^2(\vc{a},\vc{m}) \Big],
\label{eq:posteriorML}
\end{equation}
where $\mathcal{N}$ is an appropriate normalization factor, and the $\chi^2$ (often called logarithmic likelihood) is defined as
\begin{equation}
\chi^2(\vc{a},\vc{m})
=
\sum_{i,j}^{n_{\rm dat}}
\big(T^i(\vc{a}) - m^i\big)\,C_{ij}^{\rm dat}\,\big(T^j(\vc{a}) - m^j\big)
\,,
\label{eq:chi2def}
\end{equation}
where $T(\vc{a})$ represents the predictions of the model for the underlying law given some set of parameters $\vc{a}$, and $C^{\rm dat}$ defines the covariance matrix in the space of the data.

Several popular methods that have been developed to determine the posterior probability density $p(\vc{a}|\vc{m})$ have been found to be useful in contemporary QCD analyses of PDFs.
In the following, we will review the Markov chain Monte Carlo method including the Hamiltonian variant, as well as the nested sampling method. We will then review the Hessian and data resampling methods, which aim to approximate the Bayesian posterior.

% . . . . . . . . . . . . . . . . . . . . . . . . . . . . . . . . . . . . .
\subsection{Markov chain Monte Carlo}
\label{sss:MCMC}

Monte Carlo algorithms are a set of computational tools which rely on repeated random sampling. 
Steadily rooted in Bayesian statistics~\cite{DAgostini:1995jqe, 0034-4885-66-9-201, cowan1998statistical}, Monte Carlo algorithms consist of using the available data to generate samples of the posterior $p(\vc{a}|\vc{m})$ at a discrete set of points. 
The expectation value and variance of any observable $\mathcal{O}$ can then be expressed as
\begin{subequations}
\label{eq:MonteCarlomaster}
\begin{eqnarray}
E_{\rm Bayes}\{\mathcal{O}(\vc{a})\}
&=&
\frac{1}{n}\sum_{k=1}^{n}\mathcal{O}(\vc{a}_k)
\,,
\\ %\qquad
V_{\rm Bayes}\{\mathcal{O}(\vc{a})\}
&=&
\frac{1}{n}\sum_{k=1}^{n}\big[\mathcal{O}(\vc{a}_k) - E_{\rm Bayes}\{\mathcal{O}(\vc{a})\} \big]^2
\,,
\end{eqnarray}
\end{subequations}
where the parameters $\vc{a}_k$ are distributed according to the posterior $p(\vc{a}|\vc{m})$, and $n$ is the number of parameter sets sampled according to a given algorithm.

Markov chain Monte Carlo (MCMC) methods have existed since the development of Monte Carlo algorithms and rose to prominence in the 1990s.
MCMC allows one to bypass the normalization factor in Eq.~\eqref{eq:Bayestheorem} (the evidence $\mathcal{Z}$) entirely, which is usually an extremely difficult integral to calculate. 
MCMC algorithms make use of the Markov property in order to perform a  stochastic sampling of the posterior, meaning that any future sample will depend only on the current sample.
The method has the pitfall that it can still be computationally expensive, even after eliminating the need to calculate $\mathcal{Z}$, to the point of being prohibitive, depending on various elements.

Initially proposed in 1953, an algorithm which improved upon standard MCMC methods came into the statistical realm --- the Metropolis-Hastings algorithm \cite{Metropolis:1953am, Hastings:1970aa}.
This algorithm operates on the principle that certain iterations of the MCMC are accepted or rejected based on predetermined criteria with the effect of producing more precise results.
The Metropolis--Hastings algorithm proceeds as follows.
At each Monte Carlo time $t-1$, the next state $\vc{a}_t$ is chosen by sampling a candidate point $\vc{a}^\prime$ from a proposal distribution $\pi(\vc{a}_{t-1}|\vc{a}^\prime)$.
The candidate point is then accepted with the probability
\begin{equation}
\alpha(\vc{a}_{t-1},\vc{a}^\prime)
=
\min\left[
\, 
\frac{p(\vc{a}^\prime|\vc{m})\,\pi(\vc{a}_{t-1}|\vc{a}^\prime)}
{p(\vc{a}_{t-1}|\vc{m})\,\pi(\vc{a}^\prime|\vc{a}_{t-1})} \right]\,,
\label{eq:prob}
\end{equation}
and the Metropolis-Hastings transition kernel is thus
\begin{equation}
T(\vc{a}_{t-1}\to \vc{a}^\prime)
=
\pi(\vc{a}^\prime|\vc{a}_{t-1})\,\alpha(\vc{a}_{t-1},\vc{a}^\prime)\,.
\label{eq:MHtrans}
\end{equation}
If the new set of parameters $\vc{a}^\prime$ is accepted, the next state of the chain becomes $\vc{a}_t=\vc{a}^\prime$.
If it is rejected, the chain does not move and the point at the time $t$ is identical to the point at the previous time $t-1$: $\vc{a}_t=\vc{a}_{t-1}$.

A special case of the Metropolis-Hastings algorithm is the random walk Metropolis, for which the proposal distribution is chosen to be such that $\pi(\vc{a}^\prime|\vc{a}_{t-1}) = \pi(|\vc{a}_{t-1}-\vc{a}^\prime|)$.
The acceptance probability then reduces to $\alpha(\vc{a}_t,\vc{a}^\prime) = \min[1,p(\vc{a}^\prime|\vc{m})/p(\vc{a}_t|\vc{m})]$.
The efficiency of a Markov chain Monte Carlo algorithm can be optimized by choosing the proposal distribution to be as close as possible to the target distribution or alternatively, in the case of a random walk, by carefully setting the size of the scale parameter~\cite{Gbedo:2017eyp, deBerg2000}. 
Optimization becomes increasingly delicate if the number of parameters in $\vc{a}$ is large, as it is in a real fit of PDFs.

% . . . . . . . . . . . . . . . . . . . . . . . . . . . . . . . . . . . . .
\subsection{Hybrid Markov chain Monte Carlo}
\label{sss:HMC}

Hamiltonian (or hybrid) Markov chain Monte Carlo algorithms \cite{Duane:1987de} were originally developed in lattice field theory. 
These algorithms produce candidate proposals for the Metropolis-Hastings algorithm~\cite{Metropolis:1953am, Hastings:1970aa} (see Ref.~\cite{1206.1901} for an extensive review) and can be used to sample the {\it target} probability density, $p(\vc{a}|\vc{m})$.
Optimization problems can be circumvented in HMC algorithms by introducing for each set of parameters $\vc{a}$ a set of conjugate momenta $\vc{b}$, and associating a Hamiltonian 
    $H(\vc{a},\vc{b}) = \tfrac12\vc{b}^T M^{-1} \hat{b} + \mathcal{U}(\vc{a})$
to this joint state of {\it position} $\vc{a}$ and {\it momentum} $\vc{b}$, where $M$ is a mass matrix, generally taken to be diagonal, and $\mathcal{U}(\vc{a})$ an arbitrary potential energy.
This allows one to define a joint distribution
\begin{equation}
p(\vc{a},\vc{b})
=
\mathcal{N}\exp\big[-H(\vc{a},\vc{b})\big]
=
\mathcal{N}\exp\big[-\mathcal{K}(\vc{b})\big]\exp\big[-\mathcal{U}(\vc{a})\big]\,,
\label{eq:jointdist}
\end{equation}
where $\mathcal{N}$ is the normalization constant.

A common choice for the potential energy is 
    $\mathcal{U}(\vc{a})=-\log[p(\vc{m}|\vc{a})\,p(\vc{a})]$.
Starting from a point $\vc{a}_0$ of the chain, the HMC algorithm selects some initial momenta $\vc{b}_0$, normally distributed around zero, and proceeds to let the system evolve deterministically for some time according to Hamilton's equations of motion for $H(\vc{a},\vc{b})$.
The algorithm then reaches a candidate point $(\vc{a}_1,\vc{b}_1)$, which is accepted with probability $\alpha(\vc{a}_1,\vc{b}_1)=\min(1,e^{-\Delta H})$, where $\Delta H$ is the variation of the Hamiltonian with respect to the previous point.
In principle, the dynamics conserve energy, {\it i.e.} $\Delta H=0$, along a trajectory, so that the acceptance rate is 100\%.
This acceptance is degraded by numerical errors, since Hamilton's equations are solved numerically.
Such errors allow one to explore paths in the parameter space where $H$ is not conserved, and are therefore essential to determine the sampling.
The numerical noise can be efficiently exploited by appropriate methods to approximate the solution to a system of differential equations.
Among these, the leap-frog method (see Sect.~2.3 in Ref.~\cite{1206.1901} for details) proved to be particularly convenient --- one has to choose only the scale parameter and the number of iterations as free parameters, which are conventionally tuned, to obtain an acceptance rate of $60\%$.

% . . . . . . . . . . . . . . . . . . . . . . . . . . . . . . . . . . . . .
\subsection{Nested sampling}
\label{subsec:NS}

Nested sampling~\cite{skilling2006} allows for an efficient evaluation of the Bayesian evidence $\mathcal{Z}$ in Eq.~\eqref{eq:Bayestheorem}, and also produces a discrete set of posterior inferences.
The expectation value and variance of any observable $\mathcal{O}$ can be expressed as
\begin{subequations}
\begin{eqnarray}
E_{\rm Bayes}\{\mathcal{O}(\vc{a})\}
&=&
\sum_{k=1}^{n}p(\vc{a_k}|\vc{m})\mathcal{O}(\vc{a_k})\,,
\\ %\qquad
V_{\rm Bayes}\{\mathcal{O}(\vc{a})\}
&=&
\sum_{k=1}^{n}p(\vc{a_k}|\vc{m})
\big[
\mathcal{O}(\vc{a_k}) - E\{\mathcal{O}(\vc{a})\}
\big]^2\,.
\label{eq:NestedSamplingmaster}
\end{eqnarray}
\end{subequations}
The key ingredient in NS is to define a {\it prior volume} $X$, such that
\begin{equation}
X(\lambda)=\int_{\mathcal{L}(\vc{a};\vc{m})>\lambda} \dd{\vc{a}}\, p(\vc{a}),
\label{eq:priorvol}
\end{equation}
where the integral extends over the region(s) of parameter space contained within the iso-likelihood contour $\mathcal{L}(\vc{a})=\lambda$.
The multi-dimensional evidence integral in Eq.~\eqref{eq:Bayestheorem} can then be transformed into a simpler one-dimensional integral,
\begin{equation}
\mathcal{Z}=\int_0^1 \dd{X} \mathcal{L}(X),
\label{eq:monodimZ}
\end{equation}
where $\mathcal{L}(X)$, the inverse of Eq.~\eqref{eq:priorvol}, is a  monotonically decreasing function of $X$.
Therefore, if one can evaluate the likelihood $\mathcal{L}_i\equiv\mathcal{L}(X_i)$ over a discrete sequence of decreasing values $0<X_{n-n_{\rm act}}<\dots<X_i<\dots<X_1<X_0=1$, the evidence can be computed numerically using, {\it e.g.}, standard quadrature methods, as a weighted sum
\begin{equation}
\mathcal{Z}=\sum_{i=1}^{n-n_{\rm act}}\mathcal{L}_i\, w_i\,.
\label{eq:Zdiscrete}
\end{equation}
For instance, for the simple trapezium rule, the weights would be given by
$w_i=\frac{1}{2}(X_{i-1}-X_{i+1})$.

The summation in Eq.~\eqref{eq:Zdiscrete} is performed with an appropriate iterative algorithm as follows.
At the starting iteration ($i=0$), a flat prior $p(\vc{a})$ is assumed so that its volume is $X_0=1$, from which $n_{\rm act}$ {\it active} or {\it live} samples are drawn.
These samples are then ordered according to their increasing likelihood value.
The sample corresponding to the smallest likelihood $\mathcal{L}_0$ is removed from the active set, hence becoming {\it inactive}. 
It is then replaced by a sample drawn from the prior, requiring that its likelihood value is $\mathcal{L}>\mathcal{L}_0$.
The corresponding volume delimited by the iso-likelihood contour $\mathcal{L} > \mathcal{L}_0$ is described, at the first iteration, by a random variable $X_1=t_1X_0$, where $t_1$ is distributed according to the probability for the largest of $n_{\rm act}$ samples drawn uniformly from the interval $[0,1]$, $\text{Pr}(t)=n_{\rm act}t^{n_{\rm act}-1}$.
The procedure is repeated for each subsequent iteration $i$ until the entire prior volume has been traversed through nested likelihood shells; the sample associated with the lowest likelihood $\mathcal{L}_i$ in the active set is removed, then it is replaced with a new sample drawn from the prior with $\mathcal{L}>\mathcal{L}_i$, and finally the corresponding prior volume is reduced to $X_{i+1}=t_{i+1}X_i$.
The algorithm stops when the evidence $\mathcal{Z}$ in Eq.~\eqref{eq:Zdiscrete} is determined to some specified precision.

To be precise, an upper limit is set from the remaining active points by assuming that the largest evidence contribution made up by the residual portion of the posterior is $\Delta\mathcal{Z}_i=\mathcal{L}_{\rm max}X_i$, where $\mathcal{L}_{\rm max}$ is the maximum likelihood in the set of active points at the last iteration.
The algorithm stops when this quantity no longer changes the final evidence estimate, {\it i.e.}, $\Delta\mathcal{Z}_i<\epsilon$, with $\epsilon$ arbitrarily small.
Finally, an increment from the set of $n_{\rm act}$ points is added to 
the evidence estimate, Eq.~\eqref{eq:Zdiscrete},
\begin{equation}
\mathcal{Z}^\prime=\mathcal{Z}+\Delta\mathcal{Z}\,,
\qquad
\Delta\mathcal{Z}=\sum_{j=1}^{n_{\rm act}}\mathcal{L}_j w_{n-n_{\rm act}+j}\,, 
\label{eq:Zprime}
\end{equation}
where $w_{n-n_{\rm act}+j}=X_{n-n_{\rm act}}/n$ for all $j$.
Once the evidence $\mathcal{Z}$ is found, the posterior can be determined using the full sequence of active and inactive points generated in the nested sampling process as
\begin{equation}
p(\vc{a_k}|\vc{m})
=
\frac{\mathcal{L}_k w_k}{\mathcal{Z}^\prime}\,,
\qquad
k=1,\dots,n\,.
\label{eq:NSposteriors}
\end{equation}
The posterior can then be used to compute the expectation value and variance in Eq.~\eqref{eq:NestedSamplingmaster}.

%..........................................................................
\subsection{The Hessian method} 

The Hessian method was first developed in Refs.~\cite{Pumplin:2001ct, Martin:2002aw}, and essentially attempts to approximate the posterior distribution as a multi-variate Gaussian distribution in parameter space. 
To do this, one first finds the set of parameters $\vc{a}_0$ corresponding to the maximum {\it a~posteriori} (MAP) estimate, which can be obtained by minimizing the $\chi^2$ function assuming flat priors.
In a neighborhood of $\vc{a}_0$, the $\chi^2(\vc{a})$ function can be expanded up to the quadratic term in its parameters and the posterior approximated by a multi-dimensional Gaussian function,
\begin{align}
    p(\vc{a}|\vc{m})
    \propto \exp\Big( -\frac{1}{2}\chi^2(\vc{a},m) \Big)
    \propto \exp\Big( -\frac{1}{2}\Delta \vc{a}^T\, H\, \Delta \vc{a} \Big).  
\end{align}
In the last term, $\Delta\vc{a} = \vc{a}-\vc{a}_0$, the Hessian matrix elements are given by 
\begin{equation}
H_{ij} 
= \frac12
\left.
\frac{\partial^2\chi^2(\vc{a})}{\partial a^i \partial a^j}
\right|_{\vc a=\vc{a}_0 }\,,
\quad
i,j=1,\dots n_{\rm par}\,.
\label{eq:hessmatrix}
\end{equation} 
and $\exp(-\frac12\chi^2(\vc{a}_0))$ has been absorbed into the normalization of the posterior.

The approximated posterior can be reparametrized in terms of the eigendirections of the Hessian matrix via
\begin{equation}
\vc{a}(\vc{t})=\vc{a}_0+\sum_{k=1}^{n_{\rm par}} t_k\frac{\vc{e}_k}{\sqrt{w_k}}\,,
\label{eq:parvar}
\end{equation}
where $\vc{e}_k$ and $w_k$ are the orthonormal eigenvectors and eigenvalues of the Hessian matrix, respectively.
The parametrization \eqref{eq:parvar} defines a linear change of variables for the posterior, %\textit{i.e.}, 
    $p(\vc{a}|\vc{m}) \to p(\vc{t}|\vc{m})$, 
with $t_k$ parametrizing straight paths in the parameter space that radiate from $\vc{a}_0$ and run parallel to the corresponding eigenvector.
In the new space, the approximated posterior explicitly factorizes along each eigendirection,
\begin{align}
    p(\vc{t}|\vc{m})\, =\, \prod_k p_k(t_k|\vc{m})\
    \propto\ \prod_k \exp\Big(-\frac12 t_k^2\Big)\; ,
\label{eq:fact}
\end{align}
and allows one to estimate the expectation value of an observable in a simple way,
\begin{align}
E\{\mathcal{O}(\vc{a})\} 
    = \int\!\dd^n{t} ~p(\vc{t}|\vc{m})\, \mathcal{O}(\vc{a}(\vc{t}))\, 
    \approx\, \mathcal{O}(\vc{a}_0) \; .
\label{eq:expect}
\end{align}
The observable's variance can, in turn, be calculated as
\begin{align}
V\{\mathcal{O}(\vc{a})\}
    &= \int \dd^n{t} ~p(\vc{t}|\vc{m})
      \big[ \mathcal{O}(\vc{a}(\vc{t})) - E\{\mathcal{O}(\vc{a})\} \big]^2
      \notag\\
    &\approx \prod_k \int \dd{t_k} ~p\Big(t_k \frac{\vc{e}_k}{\sqrt{w_k}}\Big|\vc{m} \Big)
        \bigg( \sum_l\left.\frac{\partial \mathcal{O}\left(\vc{a}(\vc{t})\right)}{\partial t_l}
        \right|_{\vc{a}_0}
        t_l
        \bigg)^2
      \notag\\
    &=  \sum_k T_k^2\,
        \bigg(\left.\frac{\partial \mathcal{O}\left(\vc{a}(\vc{t})\right)}{\partial t_k}
        \right|_{\vc{a}_0}
        \bigg)^2
        \; ,
\label{eq:t-expansion}
\end{align}
where in the second line $\mathcal{O}$ is expanded to first order in $\vc{t}$, and $T_k$ in the third line is the \textit{tolerance factor}, defined by
\begin{align}
T_k^2= \int \dd{t_k}\, p_k(t_k|\vc{m})\, t_k^2.
\label{eq:tolerance}
\end{align}
Whenever the posterior is approximately Gaussian, one expects $T_k \approx 1$.
Introducing a finite step $\xi_k$ in each eigendirection, the variance \eqref{eq:t-expansion} can then be calculated numerically as  
\begin{align} 
V\{\mathcal{O}(\vc{a})\}
    & \approx \sum_k %\frac{1}{4}
        \frac{T_k^2}{4\xi_k^2}
        \Big[
        \mathcal{O}\Big(\vc{a}_0+\xi_k\frac{\vc{e}_k}{\sqrt{w_k}}\Big)
        -\mathcal{O}\Big(\vc{a}_0-\xi_k\frac{\vc{e}_k}{\sqrt{w_k}}\Big)
        \Big]^2\; .
\end{align}
The step size $\xi_k$ can be chosen arbitrarily, as long as it is sufficiently small to guarantee the linear approximation in $\vc{t}$. 
However, it is convenient to set $\xi_k=T_k$ in order to write the variance as 
\begin{align}
V\{\mathcal{O}(\vc{a})\}
    &\approx \sum_k\frac{1}{4}
        \Big[
        \mathcal{O}\Big(\vc{a}_0+T_k\frac{\vc{e}_k}{\sqrt{w_k}}\Big)
        -\mathcal{O}\Big(\vc{a}_0-T_k\frac{\vc{e}_k}{\sqrt{w_k}}\Big)
        \Big]^2 \; .
\label{eq:var}
\end{align}

Note that the expressions \eqref{eq:expect} and  \eqref{eq:var} for the statistical estimators have been derived under three assumptions: 
(i) the quadratic approximation of the $\chi^2$ function, producing a Gaussian posterior;
(ii) the ensuing factorization of the posterior along each eigendirection;
and
(iii) the linear approximation of $\mathcal{O}(\vc{a}(\vc{t}))$ in $\vc{t}$.
In practice, however, the Guassian approximation for the posterior is not always valid in all eigendirections, and one of the common symptoms is that the value of $T_k$ deviates from unity for eigendirections with very small eigenvalues.
Such a situation that can arise because of a lack of experimental constraints, even when the data set being fitted is statistically consistent.
To address this issue, we propose to evaluate Eq.~\eqref{eq:tolerance} using the unapproximated posterior as  
\begin{align}
T_k^2=
\int \dd{t_k}\, \frac{1}{Z_k}\exp\bigg[-\frac{1}{2}\chi^2 \big(t_k \vc{e_k}/\sqrt{w_k},m\big)\bigg]\, t_k^2\,,
\label{eq:tolerance2}
\end{align}
where the normalization $Z_k$ can be efficiently estimated numerically since it only requires a one dimensional integration over the variable $t_k$.
As long as the posterior remains approximately factorized for $t_k$ values of $O(T_k)$, using the tolerances \eqref{eq:tolerance2} instead of assuming them all equal to 1 provides a faithful representation of an observable's variance, even when the Gaussian approximation for the posterior breaks down.

In actual global QCD analyses of PDFs, one also needs to accommodate statistical inconsistencies among data sets, which cannot be accounted for by the $\chi^2$-based likelihood \eqref{eq:posteriorML}.
Indeed, such a likelihood is multiplicative with respect to each data set,     ${\cal L} = \prod_{\rm set} {\cal L}_{\rm set}$, 
and does not properly account for a scenario in which two or more data sets are mutually incompatible.
To alleviate this problem without adopting a different likelihood, a procedure known as the {\it tolerance criterion} has been advocated~\cite{Martin:2009iq, Harland-Lang:2014zoa, deFlorian:2014xna, deFlorian:2017lwf}, where the tolerance factors in Eq.~\eqref{eq:tolerance} are inflated by demanding that a given fraction of the fitted data falls within the variance of the corresponding observable.
Values %${\wm T_k^{\wm\text{\,t.c.}}
$T_k \sim 5-10$~\cite{Gao:2017yyd} are typically needed to account for 68\% of the data in PDF global analyses.

The tolerances can also be adjusted by requiring that the $\chi^2$ increases above the minimum by a specified $\Delta\chi^2$ value, namely by solving $\chi^2\big(T_k\vc{e}_k/\sqrt{w_k},m\big) = \Delta\chi^2$ \cite{Accardi:2021ysh}.
With $\Delta \chi^2 = 1$, one should obtain tolerances comparable to Eq.~\eqref{eq:tolerance2}, and a comparison to the tolerance criterion is possible by setting $\Delta\chi^2 = T_k$.
In our analysis we will perform fits to a statistically consistent set of mock data, and therefore will not need to invoke a tolerance criterion.
Nonetheless, the tolerance will need to be explicitly calculated according to Eq.~\eqref{eq:tolerance2} to accommodate for genuine deviations from the Gaussian behavior and to obtain accurate variance estimates.

% . . . . . . . . . . . . . . . . . . . . . . . . . . . . . . . . . . . . .
\subsection{Data resampling}
\label{subsec:MC}

The data resampling method for PDFs was originally pioneered by the NNPDF collaboration some twenty years ago~\cite{Forte:2002fg, DelDebbio:2004xtd} and has been adopted more recently by the JAM collaboration~\cite{Sato:2016tuz, Sato:2016wqj, Ethier:2017zbq}. 
Data resampling allows one to approximate the Bayesian posterior distribution through the use of frequentist statistics. By contrast to the methods discussed so far, frequentist methods do not make use of Bayes' theorem at all.
For a frequentist, there exists a definite true set of parameters $\vc{a}_{\true}$, and the job of frequentist parameter inference is to obtain best estimates of these unknown true values. 
The frequentist probability of an observable taking on a certain value is simply equal to the number of times that value is observed out of a number of repeatable trials.
As such, the only information the frequentist has access to is the likelihood $\mathcal{L}(\vc{a};\vc{m}$). 
It is common to choose the estimate of the true parameter values to be the set of parameters that maximizes the likelihood --- the so-called maximum likelihood estimate (MLE).

Various algorithms can be used to perform this maximization, however, the main point of interest here is how the uncertainty would be characterized from a frequentist perspective.
In the Bayesian approach one would have access to the posterior distribution, and could therefore generate credible intervals which describe the degree of belief about a parameter taking on a certain value. 
The frequentist approach must instead construct confidence intervals, in which the true observables are expected to fall as a proportion of a total number of trials.

One might wonder why one would bother to introduce alternatives to the Bayesian methods described previously.
The answer is that Bayesian techniques tend to require many likelihood evaluations in order to build up sufficient samples to gain a meaningful picture of the posterior, to the point of being ineffective in the context of real PDF data with many free parameters to estimate and large data sets.
Since the use of frequentist statistics involves finding the MLE, one often requires far fewer evaluations of the likelihood in order to find a good approximation to the location of the maximum likelihood, depending on the maximization algorithm being used.

Data resampling makes use of the traditional method of generating confidence intervals within frequentist statistics, which involves repeat experimentation.
Rather than performing a single likelihood maximization on the mean value of the data, $n_{\rm rep}$ pseudodata replicas of the original measurements $\vc{m}$ are generated following a multidimensional Gaussian smearing with the mean and uncertainties of the original measurements. 
This essentially creates $n_{\rm rep}$ new experimental datasets on which likelihood maximization can be performed, resulting in an equal number of replica parameter sets $\vc{a}_{\rm rep}$.
Distributions of the MLEs in the space of the parameters can then be obtained, from which we can calculate the mean and standard deviation for the corresponding observables. 
Since the frequentist approach does not have direct access to the posterior, Eqs.~(\ref{eq:EVbayes}) cannot be applied. 
Instead, data resampling makes use of the equations
\begin{subequations}
\begin{eqnarray}
E_{\rm freq}\{\mathcal{O}(\vc{a})\}
&=&
\frac{1}{n_{\rm rep}}\sum^{n_{\rm rep}} \mathcal{O}(\vc{a}_{\rm rep})
\,,
\\ %\qquad
V_{\rm freq}\{\mathcal{O}(\vc{a})\}
&=&
\frac{1}{n_{\rm rep}}\sum^{n_{\rm rep}}
\big[ \mathcal{O}(\vc{a}_{\rm rep}) - E_{\rm freq}\{\mathcal{O}(\vc{a})\} \big]^2.
\label{eq:Frequentistmaster}
\end{eqnarray}
\end{subequations}
Although these equations closely resemble Eqs.~\eqref{eq:MonteCarlomaster}, the quantities that are being summed over in this case are frequentist MLEs rather than Bayesian MC samples. 
In JAM global analyses the interpolant is provided by functions of the form $x^{\alpha} (1-x)^{\beta} P(x)$, where $P(x)$ is a low order polynomial, while in NNPDF analyses the function $x^{\alpha} (1-x)^{\beta}$ acts as a ``preprocessing factor'' designed to speed up the minimization and is weighted by $P(x) \to$ a~neural network.

Importantly, the distribution of the frequentist maximum likelihood estimators can be shown to coincide with the Bayesian posterior distribution given certain assumptions (namely, flat priors and a Gaussian likelihood), which implies that $E_{\rm Bayes} \approx E_{\rm freq}$ and $V_{\rm Bayes} \approx V_{\rm freq}$.
One such demonstration was performed in Ref.~\cite{DelDebbio:2021whr}, where it was found that distributions in the replicas obtained using the NNPDF4.0 PDFs approximately reproduced the expected Bayesian posterior distribution.
As such, it is not necessary to embrace the fully-fledged frequentist methodology in order to accept the data resampling approach.
One can argue that the data resampling method is a trick used to calculate the posterior more efficiently, albeit with some caveats.
Nevertheless, the data resampling approach can at most provide an approximation to the Bayesian posterior, and so should not be identified fundamentally as a Bayesian technique.

%%%%%%%%%%%%%%%%%%%%%%%%%%%%%%%%%%%%%%%%%%%%%%%%%%%%%%%%%%%%%%%%%%%%%%%%%%
\section{Comparing methodologies}
\label{sec:comparison}

In the following we perform a systematic comparison of the uncertainties that result from applying each of the methods discussed in Sec.~\ref{sec:formalism}.
Our investigation proceeds by fitting a set of pseudodata that has been generated according to a known underlying law and statistical noise.
We discuss the generation of the pseudodata in Sec.~\ref{subsec:datagen}, 
and the results of our fits in Sec.~\ref{subsec:results}.

% ...................................................................... 
\subsection{Construction of toy data}
\label{subsec:datagen}

\begin{figure}[t]
\centering
\includegraphics[scale = 0.45]{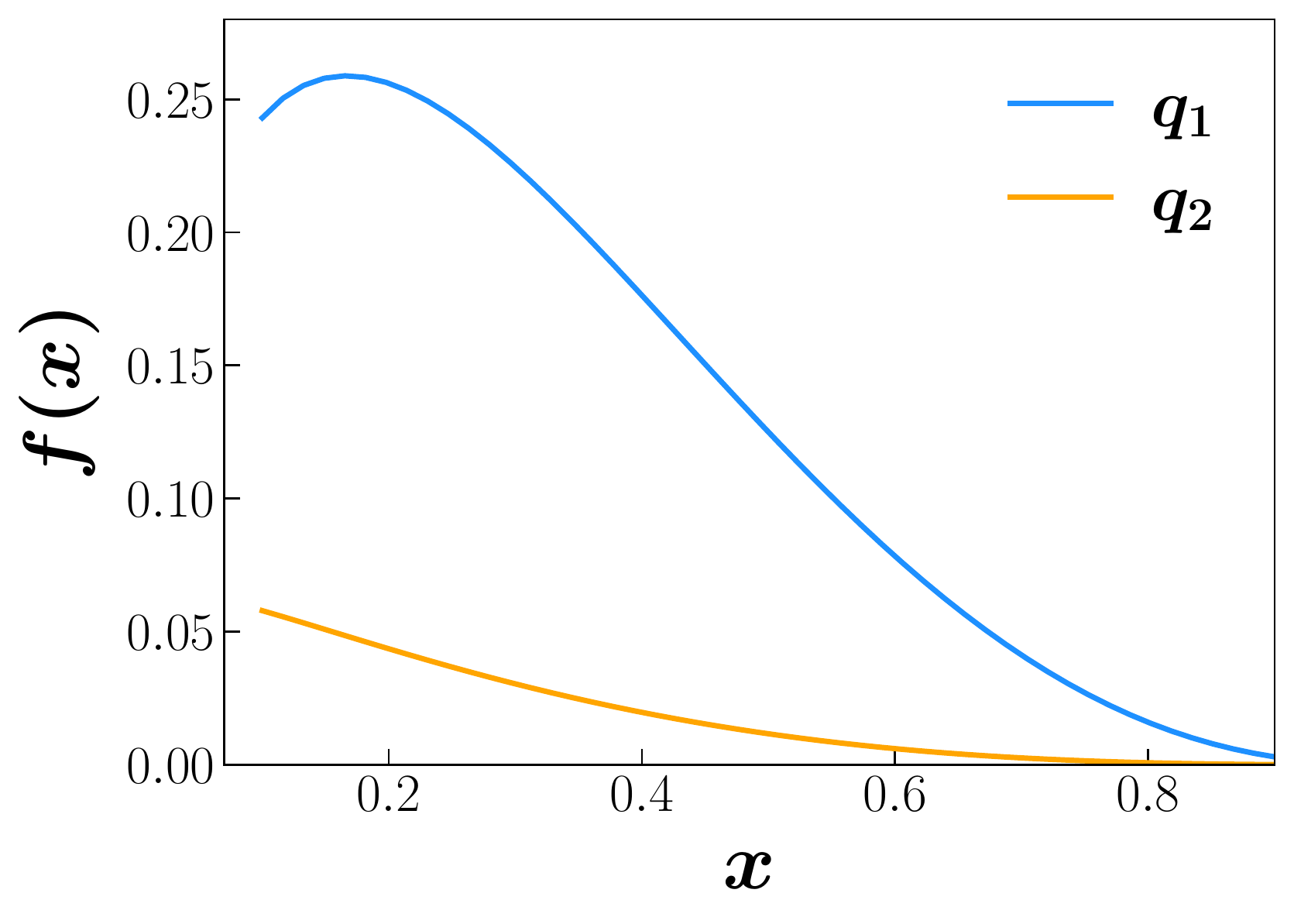}
\includegraphics[scale = 0.45]{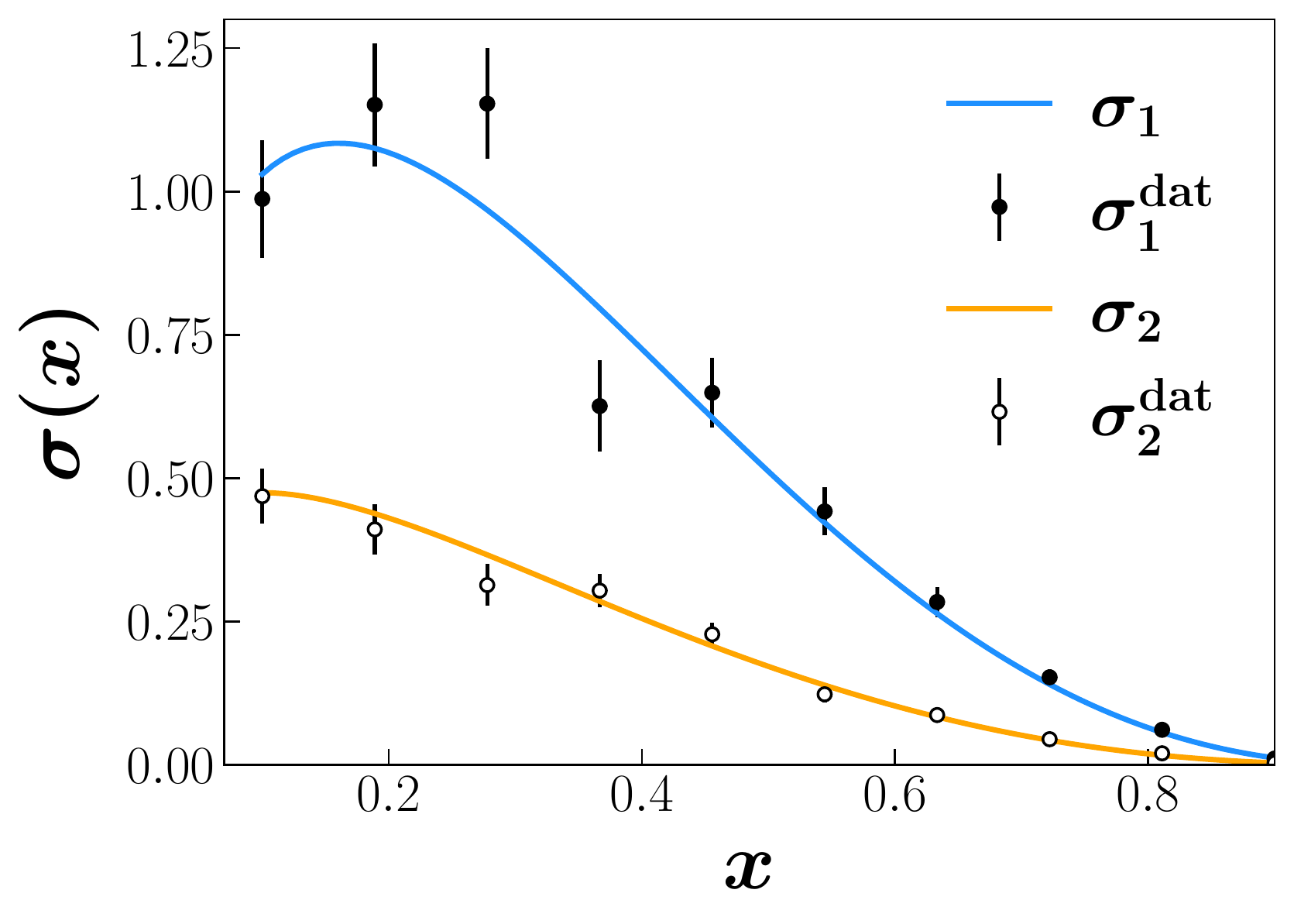}
\caption{{\it (Left)} Quark distributions $q_1$ and $q_2$ in our toy model as a function of $x$. {\it (Right)} Example of toy ``cross section'' datasets $\sigma_1$ and $\sigma_2$ for the pseudodata (filled and open circles) and cross sections calculated from the input $q_1$ and $q_2$ quark distributions (blue and orange curves).}
\label{fig:toymodel}
\end{figure}

For simplicity, we construct a set of two toy ``quark'' PDFs with momentum fraction $x$ dependence parametrized by a basic functional form,
\begin{equation}
    \begin{aligned}
     q_i(x) = N_i\, x^{\alpha_i} (1-x)^{\beta_i},\ \ \ \ i=1,2.
    \end{aligned}
    \label{eq:toymodel}
\end{equation}
These functions could, for example, represent the $u$ and $d$ quark PDFs in the proton, with free parameters $\alpha_i$ and $\beta_i$.
In this case, we can assign true values to the free parameters (which we choose to be 
$\alpha_1 = 0.5$, $\beta_1 = 2.5$, $\alpha_2 = 0.1$, $\beta_2 = 3.0$),
resulting in the functional form illustrated in Fig.~\ref{fig:toymodel}.
We can then model toy ``cross sections'' $\sigma_j$ using a linear combination of the two PDFs,
\begin{equation}
    \begin{aligned}
    \sigma_j = \sum_{i=1,2} c_{ji}\, q_i.
    \end{aligned}
    \label{eq:toysigma}
\end{equation}
These could, for instance, correspond to inclusive DIS cross sections for a proton and a neutron, with the coefficients $c_{ij}$ proportional to the squares of the quark charges, in which case we would assign
    $c_{11} = 4 c_{12} = 4 c_{21} = c_{22}$.

The values of $x$ are taken in the range $x=0.1-0.9$ at regular intervals, depending on the number of toy data points we wish to generate. 
For each $x$ value we calculate the corresponding $q_1$ and $q_2$ PDFs, from which the true $\sigma_1$ and $\sigma_2$ are determined. 
Toy cross section data points are then generated by drawing randomly from a Gaussian distribution centred on the true cross section values, and uncertainties are assigned to be 0.1 times the magnitude of each toy data point. 
An example of a toy cross section dataset with 10 points is shown in Fig.~\ref{fig:toymodel}.

% ...................................................................... 
\subsection{Fit results}
\label{subsec:results}

We perform a fit to a single dataset using each of the methods described
in Sec.~\ref{sec:formalism}, namely,  Hessian, DR, HMC and NS.
For DR we follow the traditional $x^{\alpha} (1-x)^{\beta}$ parametrization, as used by the JAM collaboration, and compare the neural network parametrization separately in Sec.~\ref{sec:bias} below.
All the methods discussed here involve parameter estimation, in which we obtain a set of Monte Carlo samples of the parameters of the model that will ultimately be used to predict the values of the cross sections and the PDFs. 
The model with which we provide each algorithm is the same as the true model from which the data are drawn (namely, Eq.~(\ref{eq:toymodel})), so that any deviation in the results from the true values is entirely due to the algorithm rather than any underlying inaccuracy in the parametrization.

For all the methods considered, bounds on the parameters were provided as $\alpha_1 \in [-1,1]$, $\beta_1 \in [0,5]$, $\alpha_2 \in [-1,1]$, and $\beta_2 \in [0,5]$, and all algorithms started in a random location in this parameter space. 
For the DR method, the \verb|SciPy minimize| \verb|least_squares| optimizer was used to minimize the $\chi^2$ for each data shuffling step, generating $10^5$ replicas.

For the Hessian method, the single $\chi^2$ minimization required to find the MAP estimator was also performed using the \verb|SciPy minimize| \verb|least_squares| optimizer. 
The $\chi^2$ itself is given by the sum of square residuals across both the $\sigma_1$ and $\sigma_2$ cross sections,
\begin{equation}
    \chi^2 = \sum_{i = 1,2}\sum_j^{n_{\rm data}}
    \left( \frac{\sigma^{\rm data}_i(x_j) - \sigma^{\rm model}_i(x_j,\vc{p})}
                {\Delta\sigma_i(x_j)}
    \right)^2,
    \label{eq:chi2min}
\end{equation}
where $i$ is the index of the two cross sections, $j$ is the index of the data points summing to a total number of points $n_{\rm data}$ in the sample, $\sigma^{\rm data}_i$ is the toy cross section data with uncertainty $\Delta\sigma_i$, and $\sigma^{\rm model}_i$ is the model cross section [Eq.~(\ref{eq:toysigma})] for a given set of parameters~$\vc{p}$. 
The Hessian is diagonalized, and the tolerances $T_k$ are calculated along each eigendirection according to Eq.~\eqref{eq:tolerance2}. 
The resulting set of values $\{T_k\} = \{1.00,1.00,1.01,1.01\}$ slightly deviates from unity for the least constrained parameter combinations, reflecting a sufficiently constraining data set for our simple toy model.
On the other hand, we have observed differences of order 10\% in the determination of the PDF uncertainties at large $x$ compared to the $T_k = 1$ Gaussian approximation. 
In real applications with many parton flavors, the deviations from 1 can be substantial in the least constrained corners of parameter space, typically leading to an overestimate of the corresponding PDF uncertainties \cite{Accardi:2021ysh}.

For HMC we used the \verb|NUTS| (No U-Turn Sampler) algorithm \cite{hoffman2011nouturn}, with the initial step size set to the default value of 0.2.
Convergence is achieved when the step size averaged across all iterations approaches the initial step size, which was found to occur with a burn-in of around $10^4$ points as well as $10^4$ adaptive points.
Finally, for the NS method we used \verb|PyMultiNest|~\cite{Feroz:2008xx}, with convergence achieved when the ratio of the largest live point contribution to the dead point integral is less than $10^{-3}$, which was found to occur with 1000 live points running for around $5\times 10^4$ iterations.

%--------------------------------------------------------------------------
\begin{figure}[t]
\centering
\includegraphics[scale = 0.5]{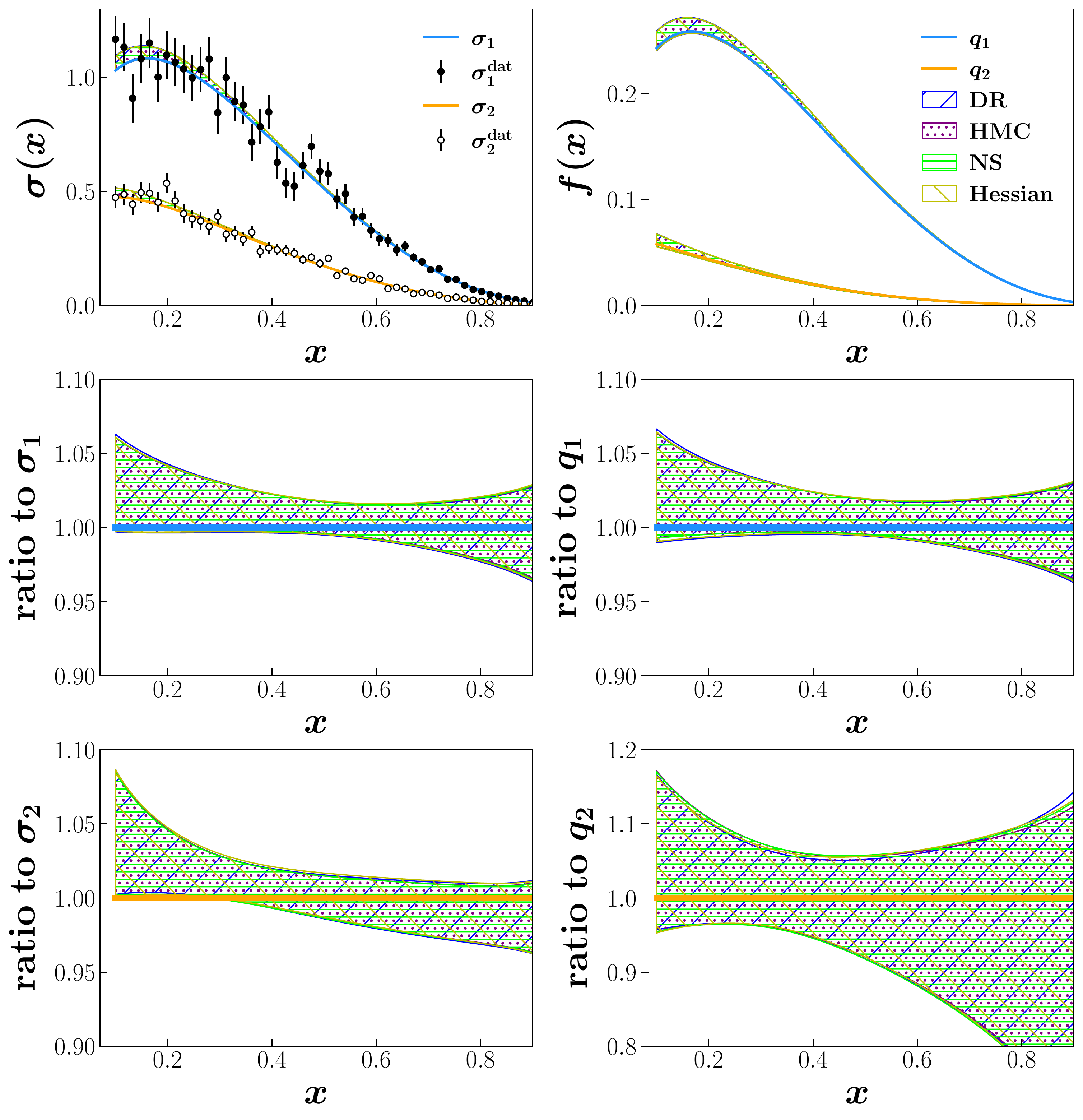}
\caption{Comparison of fit results for the DR, HMC, NS and Hessian methods. The top row contains cross section and PDF distributions, with the middle and bottom rows show ratios to the true values for each distribution.}
\label{fig:methodcomparison}
\end{figure}
%--------------------------------------------------------------------------

The predicted cross sections and PDFs for each method are shown in Fig.~\ref{fig:methodcomparison}, along with the ratios of each prediction to the underlying law.
The central values for all methods are essentially identical, as are the shapes and sizes of the predicted uncertainty bands. 
This may intuitively be expected given that the underlying law matches the parametrization utilized in each method, as well as the simplicity of the toy model used to generate the data. 
We note, however, that the agreement between the different analysis methods is nontrivial, since there are no analytical proofs to demonstrate that the DR or Hessian algorithms agree with Markovian methods such as HMC or NS. 
The reason why they are in practice equivalent is that the quantity upon which all the algorithms operate is the likelihood function.
Provided that one utilizes the same function with all the methods [in this case ${\cal L} \sim \exp(-\chi^2/2)$], uses the same parametrization for the PDFs and, of course, analyzes the same data, the results are expected to be equivalent.

Having established that with a parametrization matching the underlying law one obtains the same uncertainty estimates independent of the choice of methodology, we are in a position to make a comparison with a neural network approach.
In particular, we will explore the effect of much greater flexibility in the parametric form, and the impact of partitioning and cross validation as a stopping criterion for the underlying iterative minimization algorithm.
As we discuss in the following, the resulting PDF uncertainty estimates will be at stark variance with those found in this section.

%%%%%%%%%%%%%%%%%%%%%%%%%%%%%%%%%%%%%%%%%%%%%%%%%%%%%%%%%%%%%%%%%%%%%%%%%%
%\section{Understanding partitioning and cross validation}
\section{Neural networks, partitioning and cross validation}
\label{sec:RPCV}

The aim of this section is to explore the effect of partitioning and cross validation on PDF uncertainties in the training of neural networks (NNs).
In Sec.~\ref{sec:CV} we will describe the role that cross validation plays in fitting data with NNs, before exploring the dependence of predicted uncertainties on the partition fraction in such a framework. 
In Sec.~\ref{sec:bias} we will then compare the NN approach to the methods covered in Sec.~\ref{sec:comparison} in terms of potential inflation of uncertainties.

% ...................................................................... 
\subsection{Neural networks with cross validation}
\label{sec:CV}

Cross validation is often an essential component of the training procedure for NNs if one wants to obtain robust results that generalize to new data sets beyond those used in the training.
NN models typically have hundreds of parameters that can produce extremely complicated functional forms.
The primary metric of a NN is the loss function, which generally depends in some way on the difference between the predictions of the model and the true values of the function that it is trying to reproduce. 
Training a network involves finding the values for the weight and bias parameters of the network that minimize the loss function. 
The danger associated with such a setup is that, given a certain set of data and enough time, an NN will simply end up reproducing the data points with which it is presented.
This is referred to as {\it overfitting}, and reflects the fact that the NN model is no longer describing the underlying law that we ultimately want to learn and use to make predictions.

For our NN fits we make use of the \verb|Tensorflow| platform via the \verb|Keras| interface.
The loss function is defined in the same way as the $\chi^2$ function in Eq.~(\ref{eq:chi2min}), except that $\sigma_i^{\rm model}$ is replaced with the predicted value of the cross section according to the NN.
Our NN architecture consists of two hidden layers with 10 neurons each, with tanh activation functions and a linear output layer, similar to that used in the NNPDF4.0 analysis~\cite{Ball:2021leu}. 
For illustrative purposes, we allow the NN training to run for a long time ($10^5$ epochs) in order to deliberately overfit the data as a first example.

In this section we only perform NN fits to predict the cross section rather than inferring the PDFs themselves. 
Our goal is to hone in on the behavior of uncertainties purely due to the introduction of NNs instead of a specific underlying parametrization, and in order to study this it will be sufficiently informative to limit our analysis to the level of the cross section. 
We also dispense with the preprocessing factor $x^{\alpha} (1-x)^{\beta}$ at this stage, although we will demonstrate in Sec.~\ref{sec:bias} that including it has no significant effect on our results.

\begin{figure}[t]
\centering
\includegraphics[scale = 0.6]{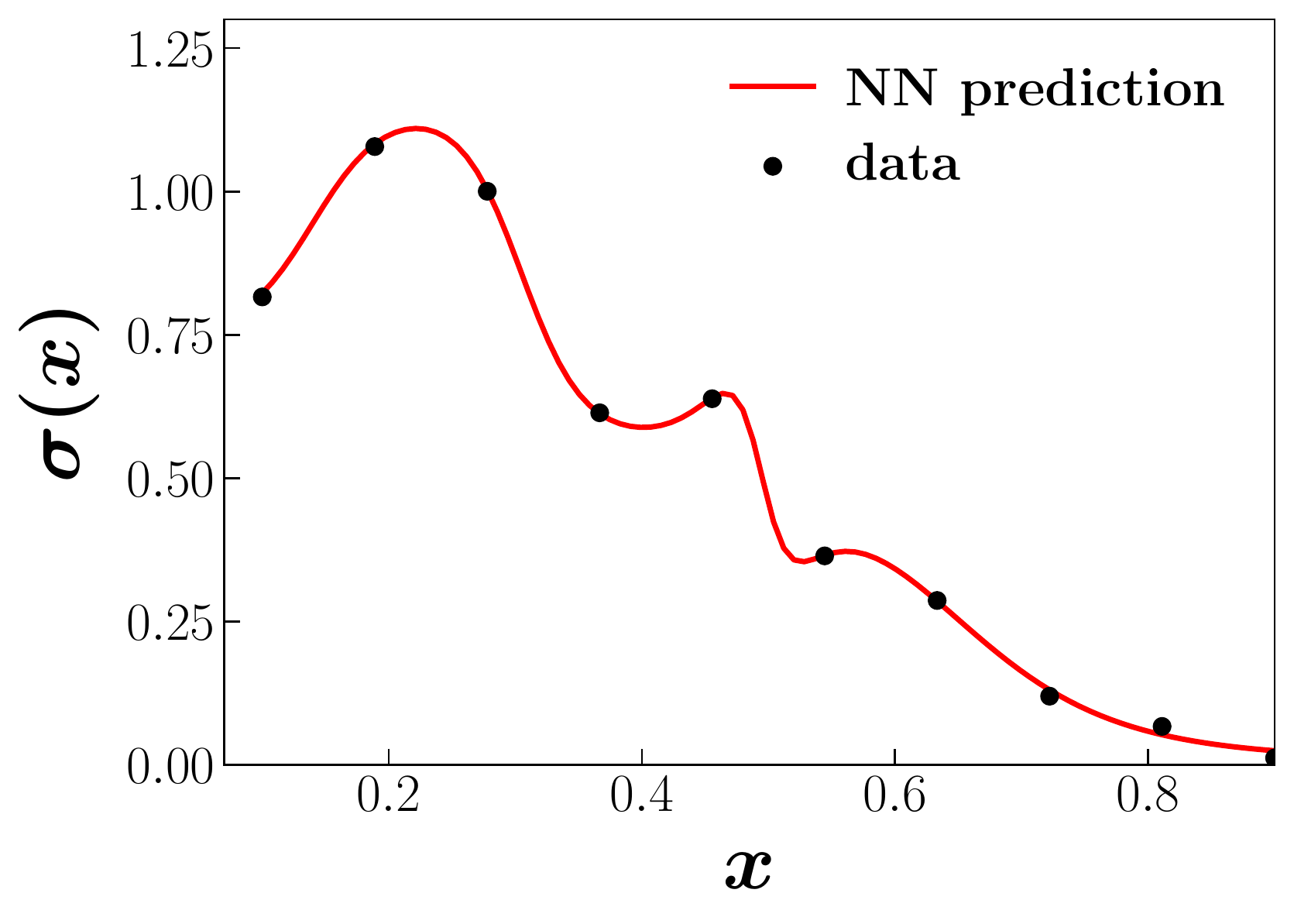}
\caption{An example of an NN overfitting a single set of our toy data.}
\label{fig:overfit}
\end{figure}

An example of overfitting our toy $\sigma_1$ data using an NN with the above settings can be seen in Fig.~\ref{fig:overfit}.
Here, the shape predicted by the NN does not resemble the true functional form at all, showing drastic and unrealistic variations in order to ensure that the data points match the prediction as closely as possible. 
Clearly this is a result we wish to avoid when applying an NN to our PDF problem.

\begin{figure}[t]
\centering
\includegraphics[scale = 0.6]{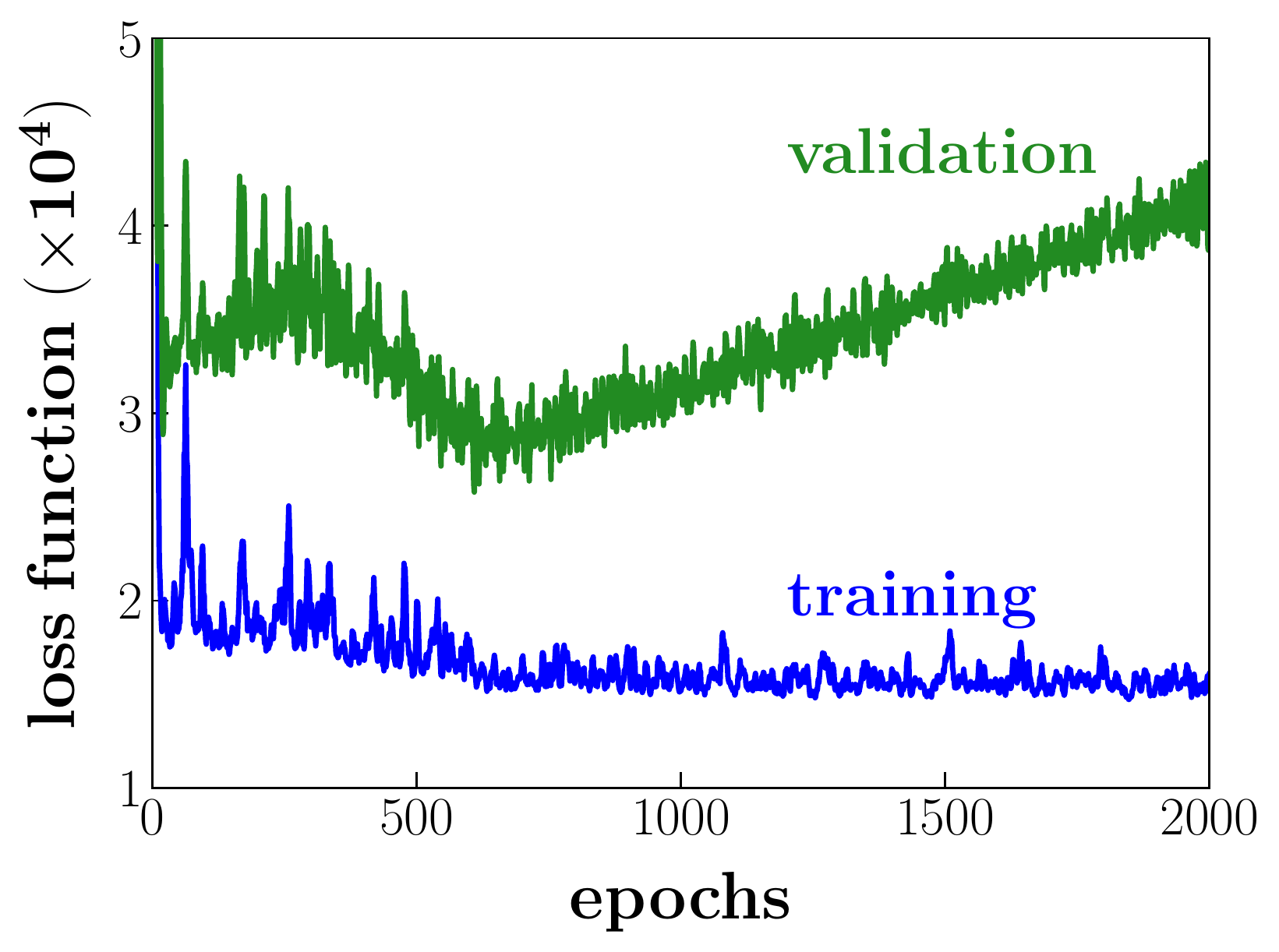}
\caption{Comparison of training (blue) and validation (green) set loss functions for an overfitting example.}
\label{fig:overfitloss}
\end{figure}

The solution to the overfitting problem is to only provide a subset of the data to the NN for it to learn, and keep the remaining data for comparison.
The proportion of the data given to the training set is called the partition fraction, $f$, and the remaining fraction $1-f$ forms the validation set.
If the loss function for the validation set does not improve during the learning process, this would be an indication that the NN is overfitting the data and the procedure should then stop.
The exact cross validation procedure used in our analysis is slightly different from this traditional early stopping criterion.
Instead of employing a stopping criterion, we allow the fit to run for a specific number of epochs (2000) and then make predictions based on the model that minimized the validation loss across all epochs, see Fig.~\ref{fig:overfitloss} for an example (here we would use the model obtained at around 600 epochs).
In this way we ensure that the best fit is found, while also avoiding overfitting the training data.

\begin{figure}[t]
\centering
\includegraphics[scale = 0.6]{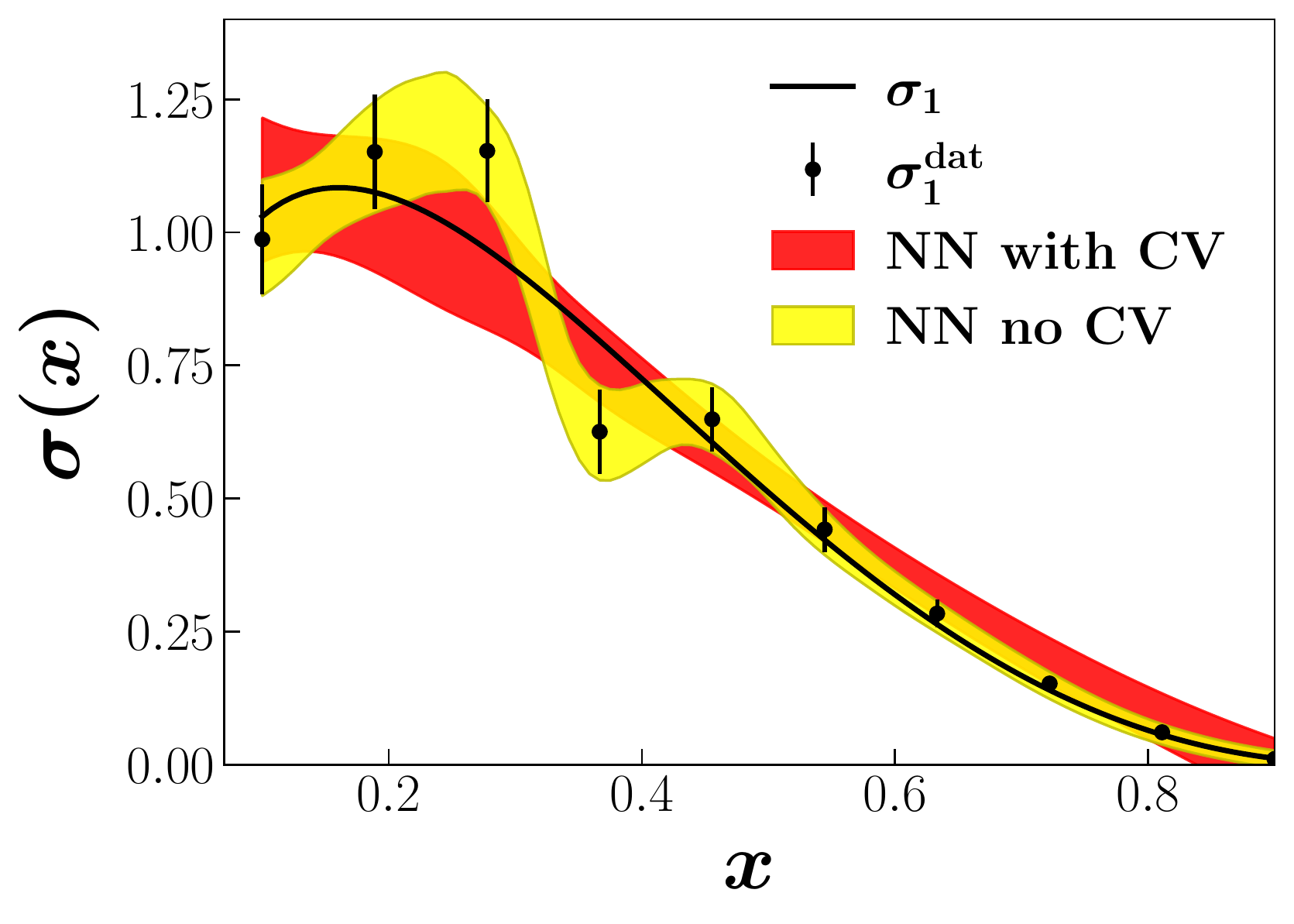}
\caption{Comparison of the $\sigma_1$ replica set for the NN fit with (red band) and without (yellow band) cross validation.}
\label{fig:sigma1_noCV}
\end{figure}

As explained in Sec.~\ref{subsec:MC}, repeated fitting of resampled data is necessary in order to construct uncertainty bands.
By comparing the predicted $\sigma_1$ distribution both with and without cross validation, as in Fig.~\ref{fig:sigma1_noCV} (for this exercise $f=0.4$), the importance of the cross validation procedure becomes clear.
Without using cross validation, the NN model simply reproduces the error bars of the data points given, and predicts an oscillatory shape that follows the data.
The NN with cross validation produces an overall smooth distribution that approximates much more closely the underlying law that we are trying to learn.

\begin{figure}[b]
\centering
\includegraphics[scale = 0.7]{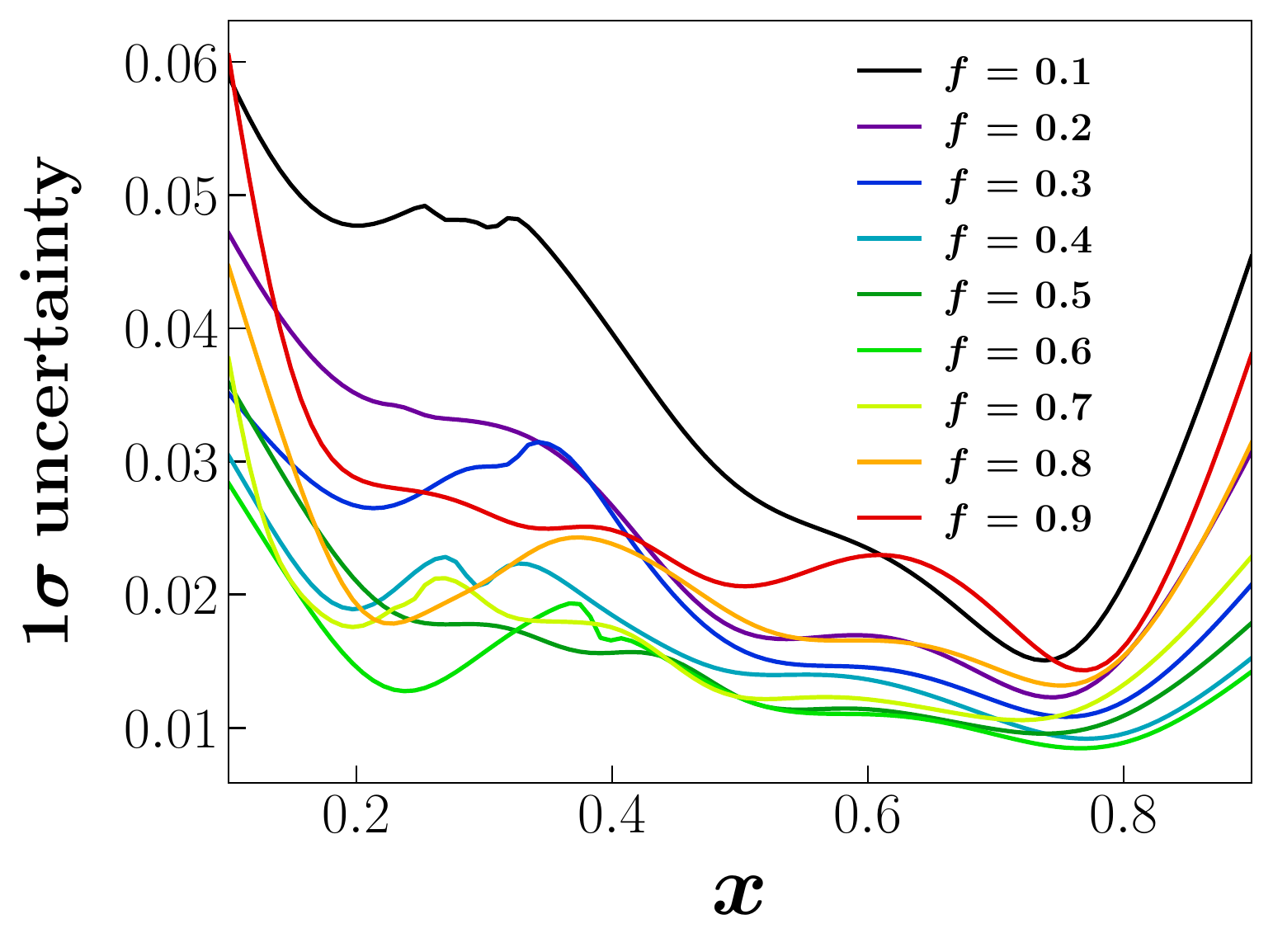}
\caption{Variation of the $1\sigma$ uncertainty with $x$ across several $f$ values from 0.1 to 0.9 for a single dataset of 50 points.}
\label{fig:stdvsf}
\end{figure}

One of the few aspects of the cross validation procedure which one can tune is the partition fraction, $f$, defining the proportion of the data on which the NN is trained.
It is interesting to consider what effect a variation in $f$ might have on the eventual uncertainty, particularly below $f=0.5$, where one might expect very little to change.
In order to study the dependence on $f$, we performed several full replica fits on the same set of toy data, varying the partition fraction between 0.1 and 0.9.
For 50 toy data points, we find that although the mean values of $\sigma_1$ are similar, the uncertainty is affected significantly.
This can be seen in Fig.~\ref{fig:stdvsf}, where the standard deviation is shown for different values of $f$.
We find that there is substantial variation in the uncertainty depending on what value of $f$ is chosen, particularly when the partition fraction is small.
In the range from $f = 0.4$ to 0.7, there are only slight differences between the uncertainties, but these are enhanced for the $f=0.1$, 0.2, 0.3, 0.8 and 0.9 curves, being up to twice as large as for $f=0.5$ for certain $x$ values.

\begin{figure}[t]
\centering
\includegraphics[scale = 0.7]{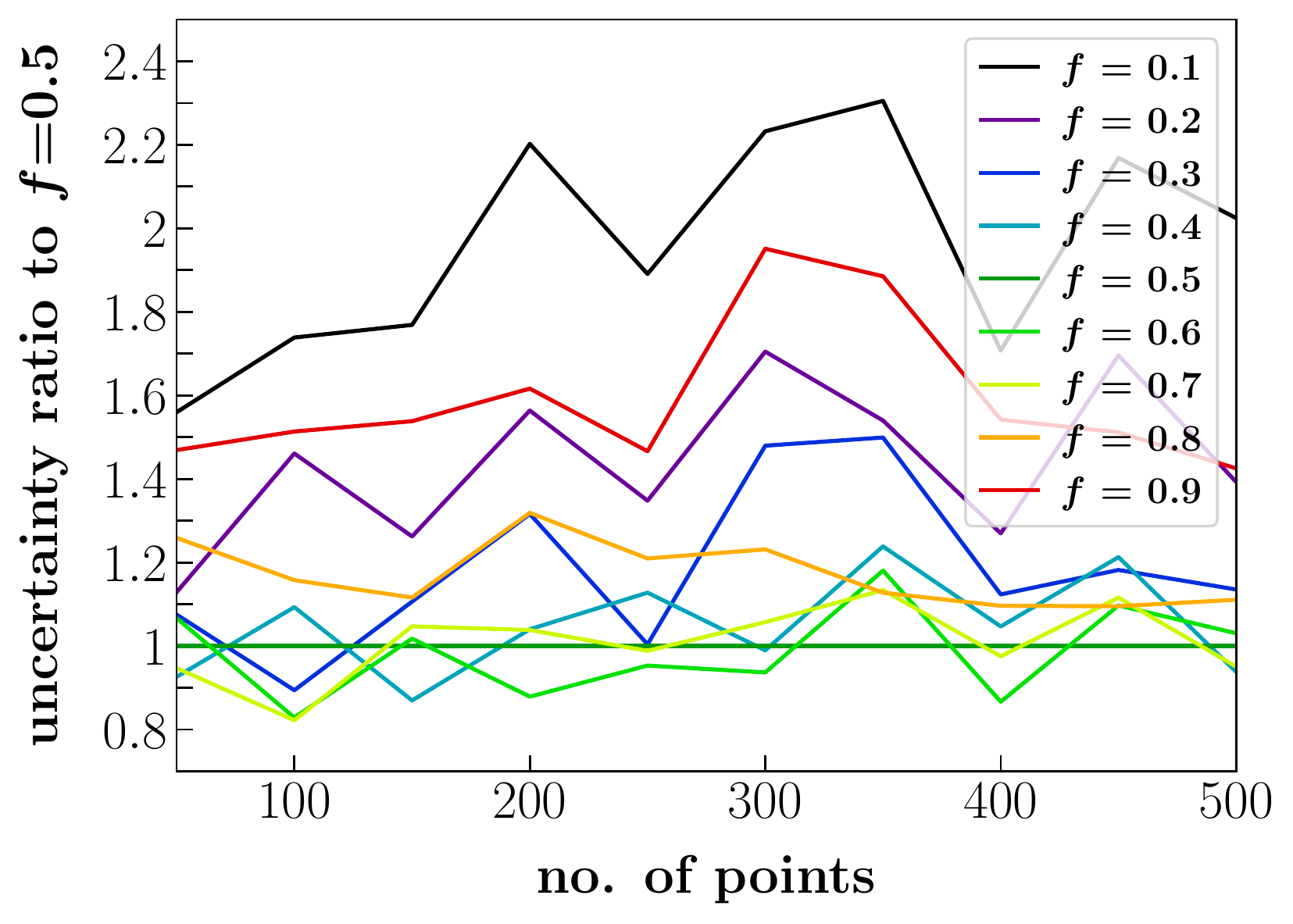}
\caption{Total uncertainty ratio to $f=0.5$ for different partition fractions as a function of number of data points.}
\label{fig:stdvsftotal}
\end{figure}

Another important question is whether the dependence of an NN fit on $f$ diminishes as the number of data points is increased.
In Fig.~\ref{fig:stdvsftotal} we increase the number of data points up to 500, plotting the total uncertainty for each value of $f$ as a ratio to the $f=0.5$ line.
This allows us to visualize the underlying trend, given that the total uncertainty for all values of $f$ should decrease as the number of points increases.
The enhancement for the $f=0.1$, 0.2 and 0.9 uncertainty seen in Fig.~\ref{fig:stdvsf} remains relatively constant as one increases the number of data points, becoming slightly more pronounced.
At the same time, the $f=0.4-0.7$ uncertainties appear consistent for all numbers of data points, with slight enhancements in total uncertainty for $f=0.3$ and $f=0.8$.
On the strength of these observations, we cannot rule out the importance of $f$ with respect to uncertainty quantification in the NN method, nor can we conclude that below $f=0.5$ an increased dataset will remove any potential artificial inflation of uncertainty due to the choice of $f$.

\newpage
%.........................................................................
\subsection{Comparing NN methods to parametric methods}
\label{sec:bias}

To further comprehend the differences between the NN methodology and that of traditional parametric modelling, it is worth comparing the results of fits to our toy data for all of the methods discussed so far. 
At this point we also include the preprocessing factor $x^{\alpha} (1-x)^{\beta}$ in the NN fits.
This is usually added with the aim of speeding up the minimization procedure. However, it is worth testing whether its inclusion could be playing a greater role in helping the NN to obtain a shape that is more consistent with the underlying law. 
We apply this factor in the same way as the NNPDF collaboration; specifically, we parametrize the PDFs as
\begin{equation}
    \begin{aligned}
     q_i(x) = x^{\alpha_i} (1-x)^{\beta_i}\, \text{NN}_i(x),\ \ \ \ i=1,2,
    \end{aligned}
    \label{eq:NNparam}
\end{equation}
where $\text{NN}_i(x)$ represents the neural network weights.
We then define effective asymptotic exponents as
\begin{equation}
\begin{aligned}
\alpha_{i}(x) = \frac{\ln{q_i(x)}}{\ln(1/x)}, \qquad\quad
\beta_{i}(x) = \frac{\ln{q_i(x)}}{\ln(1-x)}.
\end{aligned}
\label{eq:asympexp}
\end{equation}
The ranges for each of the $\alpha_i$ and $\beta_i$ values are obtained through repeated fitting by taking the envelope of twice the 68\% confidence interval for each $x$ value until the preprocessing exponents no longer change.
For each replica fit in the actual analysis, the values of $\alpha_i$ and $\beta_i$ are chosen randomly according to uniform distributions in those ranges.

\begin{figure}[t]
\centering
\includegraphics[scale = 0.5]{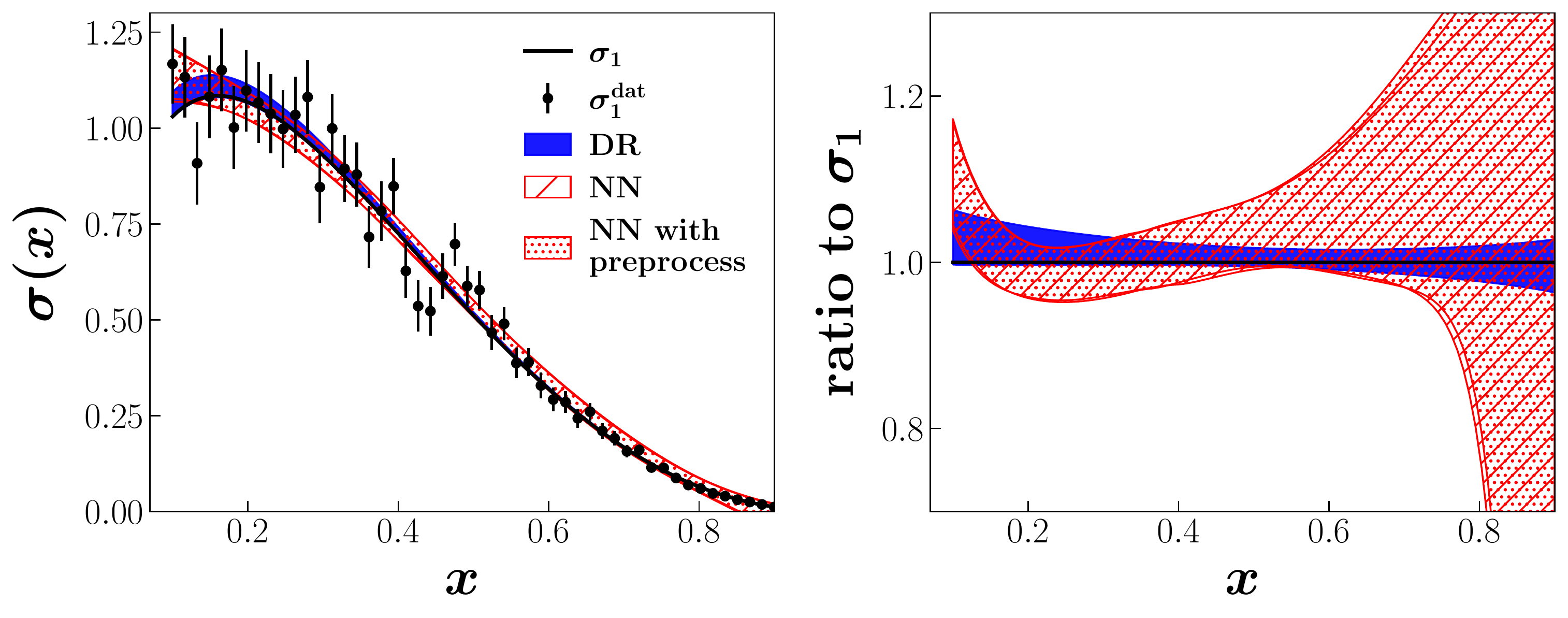}
\caption{Comparison of a single $\sigma_1$ replica set between NN and parametric DR methods (left panel), and the ratio of each predicted point to the true underlying law (right panel).}
\label{fig:singlefitcomparison}
\end{figure}

A full fit for one instance of our toy $\sigma_1$ data with 50 points is shown in Fig.~\ref{fig:singlefitcomparison}.
Here, we compare the distributions and ratio to the true values for the NN ($f$ = 0.6) with and without preprocessing, as well as parametric DR (recall that parametric DR was shown to produce the same results as the Hessian, HMC and NS).
Both NN methods produce predicted uncertainty bands that are much wider than that of parametric DR, particularly at large~$x$.
The fact that there is little difference between the NN predictions with and without preprocessing suggests that preprocessing does not play a significant role in this example.

Such a demonstration only becomes meaningful in the context of a full replica fit for many initial data sets rather than just one. 
By counting how often a given method predicts an uncertainty band that contains the true underlying law (as inspired by closure tests), we can determine whether the predicted uncertainties are statistically valid.  
Since the uncertainty bands represent the variance of the cross section around its expected value, we do not expect them to always contain the true values. 
Rather, a reasonable method will produce uncertainty bands that contain the true values around 68\% of the time across a large number of initial data sets.

\begin{figure}[t]
\centering
\includegraphics[scale = 0.7]{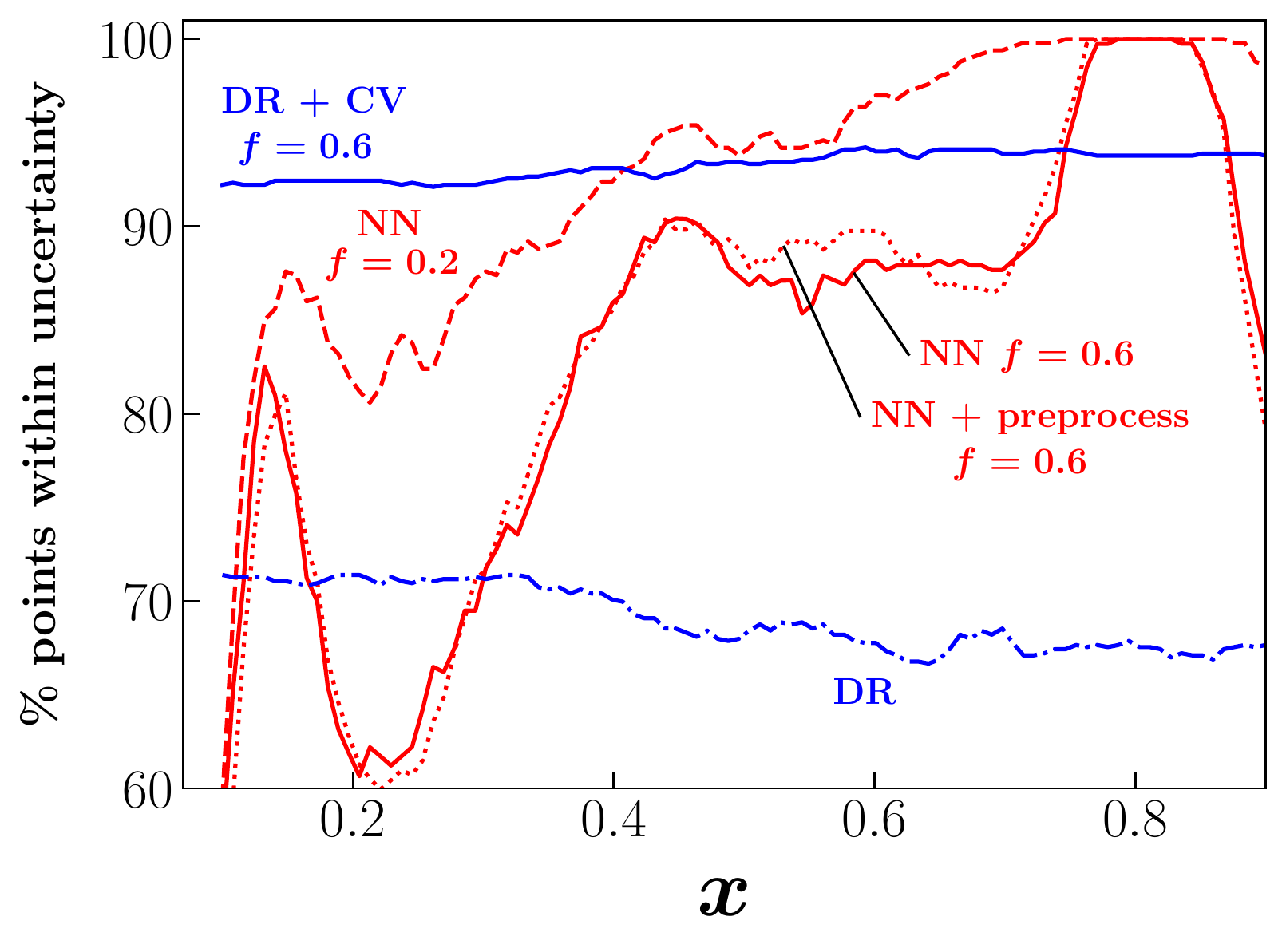}
\caption{Study of confidence interval coverage for the NN (red lines) and parametric DR (blue lines) methods as a function of $x$.} 
\label{fig:biasestimate}
\end{figure}

In Fig.~\ref{fig:biasestimate} we determine the percentage of uncertainty bands across all data sets that contain the true values for all $x$, corresponding to each method. 
We plot the NN method with $f = 0.2$ and $f = 0.6$ to again demonstrate the importance of the partition fraction, and also show the results for the NN with preprocessing.
We find that the NN predicted uncertainty bands contain the true value too often, particularly at high $x$ values.
This effect is exacerbated by reducing the partition fraction. 
By contrast, the parametric DR method consistently hovers around the 68\% value that would be expected from accurate confidence intervals.

The inflated uncertainty observed is not entirely due to the usage of neural networks in fitting the toy data.
Another contributing factor to this result is the presence of the cross validation procedure itself. In Fig.~\ref{fig:biasestimate} we also include a cross validation method in the context of parametric DR, dividing the data into a training and validation set with $f = 0.6$ and cutting off the $\chi^2$ minimization when there is no further improvement for the validation set. Comparing DR with and without cross validation, we can clearly see a substantial increase in the width of the uncertainty bands when cross validation is used, up to a similar mean value as that of NNs at the same partition fraction.

These results imply that, for our specific simplified example of PDF data, the uncertainty bands predicted by an NN fit do not realize the $1\sigma$ confidence interval that one would naively expect. 
This may be due in part to a misrepresentation of the underlying law, but also a result of the cross validation procedure that forms an integral part of any NN fit to experimental data.

One may ask how is it that NN fits, as well as partitioning and cross validation applied to simple parametric DR, can so dramatically change the variance of the fitted observable?
From our perspective, the algorithms deployed in either case effectively lead to a change in the  likelihood function.
In the case of cross validation applied to DR, this can be heuristically understood by viewing the validation $\chi^2$ as a potential that stochastically pulls the parameters away from the true minimum, thereby increasing the variance of any calculated observable, but in a uniform way.
This is illustrated with the very similar shapes for the DR and DR with CV results in Fig.~\ref{fig:biasestimate}, where one goes from an $\sim 68\%$ confidence level to an $\sim 90\%$ when using partitioning and cross validation, as if the likelihood was rescaled by the adopted stopping criterion.
NN methods not only appear to rescale the likelihood, but they also do so to different degrees across the $x$ range, resulting in a behavior that is not immediately amenable to a heuristic analysis.

%%%%%%%%%%%%%%%%%%%%%%%%%%%%%%%%%%%%%%%%%%%%%%%%%%%%%%%%%%%%%%%%%%%%%%%%%%
\section{Conclusions and outlook}
\label{sec:conclusion}

The need for reliable uncertainty quantification in global QCD analysis motivated this study to perform a systematic comparison of different methods for estimating uncertainties on PDFs and similar correlation functions.
The test laboratory employed here, in the form of the toy PDF model, although exceedingly simple, allowed us to rigorously test and verify the generation of uncertainties in a controlled context, in which the underlying physical law is known.
We were thus able to demonstrate that methods which use a parametrization that matches the underlying law lead to the same uncertainty estimates --- regardless of whether the uncertainties are determined using the traditional Hessian or data resampling methods, or explicitly Bayesian techniques such as nested sampling or HMC.
While the result may be intuitively expected, it is by no means trivial to verify in practice that all the approaches are indeed exploring the same likelihood function.

We then focused on the impact that partitioning and cross validation have on uncertainty quantification.
We found that these added elements effectively lead to a rescaling, and even a deformation of the likelihood function, which also depends on the partition fraction.
Our results suggest that this is inherent in the methodology rather than in the details of the model or the data set analyzed.

Comparing the parametric methods with NN approaches, we find indeed that NNs produce wider uncertainty bands than expected, whether or not one uses a preprocessing factor, and that the dependence on the partition fraction is not reduced as one increases the number of data points to be fitted. 
The enhanced uncertainty is found to be due, at least in part, to the use of cross validation.
While DR+CV exhibits this behavior in a uniform way in observable space, heuristically due to the additional influence that the validation data set has on the fitting of the training data set, the NN's inflation of the cross section uncertainty depends also on $x$ 
and becomes very large as $x \to 1$.
Whether this non-uniform uncertainty inflation is due to the highly flexible PDF parametrizations provided by the NNs or to other elements in the analysis algorithm remains an interesting question for future investigation.

Overall, given that the data set being analyzed by the NN and parametric methods is the same, we can only conclude that NN methods
(as well as parametric methods supplemented by cross validation) 
algorithmically deform the nominal likelihood utilized in the analysis in ways that are not necessarily 
under control. In view of these results it is natural to ask in what sense are PDF fits based on NN methods compatible with those based on Hessian + tolerance methods, since the likelihood functions that they utilize are technically different. 
From our perspective, the tolerance criterion effectively is a change in the likelihood function, which can also be cast as a rescaling of the experimental uncertainties to remove potential tensions present among the data. 
NN-based analyses posit instead that the use of a tolerance criterion is not necessary~\cite{Forte:2002fg, *DelDebbio:2004xtd, *DelDebbio:2007ee, *Ball:2008by}, but the roughly equivalent size of the uncertainties they obtain may simply be coincidental and due to the likelihood deformation and resulting uncertainty inflation that we already observed when analyzing compatible data sets in our toy model.
We thus echo the concern expressed in~\cite{Accardi:2016ndt} that a meta-analysis, such as PDF4LHC~\cite{Butterworth:2015oua}, that combines existing PDFs from different groups may not only obscure the fundamental connection between experimental data and theory, but also obfuscate the true meaning of the uncertainties, since these seem to ultimately originate from a different choice of the likelihood function.

Finally, we make a note of caution about the generality of our results, since the conclusions should be qualified by the fact that we are using a rather simplistic model. While this has been necessary in order to keep the analysis tractable and allow unambiguous conclusions to be drawn from the methodology comparisons, there may be new features present in higher dimensional analyses of more complex data sets, such as in real global QCD analyses.
We leave discussion of this issue to a future publication~\cite{HuntSmith2022}. \\

%%%%%%%%%%%%%%%%%%%%%%%%%%%%%%%%%%%%%%%%
%%%%%%%%% ACKNOWLEDGMENTS %%%%%%%%%%%%%%
%%%%%%%%%%%%%%%%%%%%%%%%%%%%%%%%%%%%%%%%
\begin{acknowledgments}

We thank E.~R.~Nocera for his collaboration on much of the material presented here.
This work was supported by the DOE contract No.~DE-AC05-06OR23177, under which Jefferson Science Associates, LLC operates Jefferson Lab; DOE contract No.~DE-SC0008791 (AA); and by the University of Adelaide and the Australian Research Council through the ARC Centre of Excellence for Dark Matter Particle Physics (CE200100008) and Discovery Projects DP180102209 (MJW) and DP180100497 (AWT).

\end{acknowledgments}

\newpage

%%%%%%%%%%%%%%%%%%%%%%%%%%%%%%%%%%%%%%%%
%%%%%%%%%   BIBLIOGRAPHY  %%%%%%%%%%%%%%
%%%%%%%%%%%%%%%%%%%%%%%%%%%%%%%%%%%%%%%%
\bibliography{bibliography}

\end{document}